\documentclass[%
preprint,
]{revtex4-2}

\usepackage{graphicx}% Include figure files
\usepackage{dcolumn}% Align table columns on decimal point
\usepackage{bm}% bold math
\usepackage{xcolor}
\definecolor{red}{RGB}{200,0,0}
\begin{document}

\title{Electronic signatures of successive itinerant, magnetic transitions in hexagonal La$_{2}$Ni$_{7}$ }

\author{Kyungchan Lee$^{1,2}$, Na Hyun Jo$^{1,2}$, Lin-Lin Wang$^{1,2}$, R. A. Ribeiro$^{1,2}$, Yevhen Kushnirenko$^{1,2}$ Ben Schrunk$^{1,2}$, Paul C. Canfield$^{1,2}$ }
 \altaffiliation[ ]{canfield@ameslab.gov}
\author{Adam Kaminski$^{1,2}$}%
 \email{adamkam@ameslab.gov}
\affiliation{$^{1}$Iowa State University, Department of Physics and Astronomy, Ames, IA, 50014}

\affiliation{$^{2}$Ames Laboratory US Department of Energy, Ames, Iowa 50011, USA}%

\date{\today}

\begin{abstract}
We use high-resolution angle-resolved photoemission spectroscopy (ARPES) and density functional theory (DFT) calculations to study the electronic and magnetic properties of La$_{2}$Ni$_{7}$, an itinerant magnetic system with a series of three magnetic transition temperatures upon cooling, which end in a weak itinerant antiferromagnetic (wAFM) ground state. Our APRES data reveal several electron and hole pockets that have hexagonal symmetry near the $\Gamma$ point. We  observe significant reconstruction of the band structure upon successive magnetic transitions at T$_{1}$ $\sim$ 61~K, T$_{2}$ $\sim$ 57~K and T$_{3}$ $\sim$42~K. The experimental data are in a reasonable agreement with DFT calculations, demonstrating their applicability  to itinerant antiferromagnet systems. Our results detail the effects of magnetic ordering on the electronic structure in a Ni-based weak antiferromagnet.

\clearpage

\end{abstract}

\maketitle

%\tableofcontents

\section{Introduction}

More than 30 years ago,  cuprate high-$T_{c}$ superconductors were reported  \cite{bednorz1986possible_cuperate_discover} and followed by discovery of iron based high-$T_{c}$ \cite{Hosono2008}. Their fascinating properties have been intensively investigated \cite{schilling1993superconductivity_tc130,ginsberg1998physical_hightc_properties,batlogg1991physical_hightc_properties, canfield2010feas_BaFe2As2, Johnston2010}.  After decades of extensive experimental and theoretical studies, a spin fluctuation scenario is the most promising mechanism for high-$T_{c}$ superconductivity\cite{miyake1986spin_spinfluc_superconduc_theory1,moriya2000spin_spinfluc_superconduc_theory2,tsuei2000pairing_SF,scalapino2012common_SF}. Consequently, weak itinerant antiferromagnets (wAFM) are particularly interesting and have drawn a lot of attention recently because of the potential for large spin fluctuations, which can lead to pairing and superconductivity \cite{moriya1990antiferromagnetic_AFM_vs_supercond_theory,moriya1991antiferromagnetic_theory,monthoux1992weak_theory,monthoux1991toward,monthoux1993yba,AFM_SC,canfield2019new_PCCadd,canfield2016preserved_PCCadd}.

Magnetic ordering in an intermetallic system can  stem from two different contributions. The the first contribution is from local magnetic moments resulting in Curie-Weiss susceptibility. The effect of local magnetic moments is well understood by the Heisenberg model in the mean-field approximation \cite{yoshioka1989mean_heisen_MFT1,gross1989ground_heisen_MFT2,heisenberg1985theorie_heisen_MFT0}. The other component is itinerant magnetism. In this regime, long range magnetic order arises due to spin fluctuations of itinerant electrons \cite{stoner1938collective_iti0,moriya1984itinerant_iti1}. Current theories work reasonably well in the limits of these two scenarios, that is, when either local magnetic moments  or itinerant fluctuations play a dominant role . 
Whereas it is relatively easy to discover or identify examples of local moment intermetallics compounds (e. g. by simply incorporating trivalent, moment bearing rare earths) the design or discovery of examples of itinerant magnetic systems is much harder given that many transition metal based intermetallic compounds simply become non-moment bearing.  Following the discovery of Fe-based superconductors\cite{Hosono2008, canfield2010feas_BaFe2As2, Johnston2010}, there has been a renewed push to design, discovery or identify itinerant, weak (or fragile) antiferromagnetic systems \cite{canfield2016preserved_PCCadd,canfield2019new_PCCadd}.

Angle resolved photo-electron spectroscopy has, recently, also been focusing on the diverse changes that magnetic ordering can have on the band structure, depending, in part on the size of the ordered moments as well as their degree of hybridization with the conduction electrons.  On one end of this spectrum is recent work on magnetic topological systems \cite{hyun2020manipulating,otrokov2019prediction_AFM_TI} as well as rare-earth bearing semi-metals such as NdBi \cite{Schrunk2022, Linlin2022, Yevhen2022}; on the other end of this spectrum is work on Fe-based superconductors such as pure and doped BaFe$_{2}$As$_{2}$~\cite{canfield2010feas_BaFe2As2,li2009superconductivity_BaFe2As2_Ni_doped,saha2010superconductivity_BaFe2As2_Pt_doped} as well as  CaKFe$_{4}$As$_{4}$\cite{mou2016enhancement_CaKFe4As4_1,biswas2017signature_CaKFe4As4_2,jost2018indication_CaKFe4As4_3}. One of the questions that has emerged is how sensitive can ARPES be to very small moment ordering.  The discovery and study of a wider range of wAFM systems will help clarify this question.

La$_{2}$Ni$_{7}$ is a rare example of a small moment ($\sim$ 0.1 $\mu_{B}$/Ni), itinerant magnetic system that shows a cascade of three magnetic ordering temperatures:  T$_{1}$ $\sim$ 61 K, T$_{2}$ $\sim$ 57 K and T$_{3}$ $\sim$ 42 K \cite{ribeiro2021small_La2Ni7_PCC} corresponding to C, B and A phases, respectively. The low temperature state has long been thought to be associated with wAFM ordering \cite{buschow1983magnetic_La2Ni7_mag1_26,parker1983magnetic_La2Ni7_mag2_27,fukase1999successive_La2Ni7_mag3_28,fukase2000itinerant_La2Ni7_mag4_29,tazuke1993magnetism_La2Ni7_mag6_32,tazuke2004metamagnetic_La2Ni7_mag5_30,ribeiro2021small_La2Ni7_PCC} and recent band structure predictions \cite{crivello2020relation_La2Ni7_DFTmain} have proposed a wAFM associated with a long wavelength  periodicity along the $c$-axis, which even more recently have been borne out by single crystal neutron diffraction measurements\cite{wilde2022}. Recent single crystal growth and characterization of this system\cite{ribeiro2021small_La2Ni7_PCC} has made it very attractive as a possible wAFM that could be amenable to ARPES measurements. Indeed the neutron scattering study \cite{wilde2022} has identified the propagating wave vectors of these three wAFM phases as all along the c-axis. While the C and B phases have the same incommensurate wave vectors, the direction of the magnetic moments are different and can be tilt away from the c-axis. In contrast, the lowest-T wAFM A phase has a commensurate wave vector of (0 0 1), which corresponds to a block-wise AFM configuration, i.e., intra-block FM but inter-block AFM coupling with the magnetic moments pointing along the c-axis, confirming the earlier proposed configuration. Each block is half of the chemical unit cell in the A state with the block-wise AFM.  

In this paper, we report  direct measurements of the band structure using the ARPES and compare them to DFT calculations for both paramagnetic (PM)  and the lowest temperature AFM state that has had it magnetic structure predicted by DFT and confirmed by neutron scattering measurements\cite{wilde2022}. We find that magnetic transitions in La$_{2}$Ni$_{7}$ are accompanied by observable changes in the band structure, both at E$_F$ and well below. We observe that the PM state of La$_{2}$Ni$_{7}$ is characterized by several electron and hole pockets near the $\Gamma$ and $M$ points and zone boundaries, in a reasonable agreement with DFT calculations. In the magnetically ordered states, several band splitting occur around the $\Gamma$  and $M$ points. While our detailed analysis of  ARPES data suggest that the successive band splittings are more pronounced in the $K$- $\Gamma$-$K$ direction, these effects are also observed along  $M$-$\Gamma$-$M$ in the proximity of $E_{F}$.

\section{methods}
Single crystals of La$_{2}$Ni$_{7}$ were grown out of a La-rich (relative to La$_{2}$Ni$_{7}$) binary, high-temperature melt. Elemental La (Ames Laboratory, 99.99+\% pure) and Ni (Alpha, 99.9+\% pure) were weighed out in a La$_{33}$Ni$_{67}$ atomic ratio and placed into a tantalum crucible which was sealed with solid caps on each end and a perforated cap in the middle to act as a filter for decanting.\cite{canfield2019new_PCCadd,canfield2001high_grow1} The assembled Ta crucible was then itself sealed into an amorphous silica tube with silica wool above and below it to act as cushioning. This growth ampoule was then place in a resistive box furnace. The furnace was then heated to 1150 $^{\circ}$C over 10 hours, held at 1150 $^{\circ}$C for 10 hours, cooled to 1020 $^{\circ}$C over 4 hours and then very slowly cooled to 820 $^{\circ}$C over 300 hours at which point the growth ampoule was removed and decanted in a centrifuge to separate the La$_{2}$Ni$_{7}$ single crystals from the residual liquid. \cite{canfield2019new_PCCadd} Crystals grew as well faceted plates with clear hexagonal morphology (see Fig. \ref{fig:fig1} (c) ).  As shown in Ref. \cite{ribeiro2021small_La2Ni7_PCC} the crystallographic c-axis is perpendicular to the hexagonal surface of the plates. Previous studies indicated that La$_{2}$Ni$_{7}$ crystallizes in two different structures (rhombohedral or hexagonal) based on stacking of the layers along the $c$ direction. The single crystals grown using  method described above were single phase with a hexagonal Ce$_{2}$Ni$_{7}$ type structure. The samples show magnetic ordering below T$_{1}$ = 61~K and underwent successive re-ordering at T$_{2}$ = 57~K and T$_{3}$ = 42~K.

The samples were cleaved \textit{in situ} at base pressure lower than 5 $\times$ 10$^{-11}$ Torr, yielding shiny, flat, mirrorlike surfaces along the basal plane of the hexagonal structure. High photon energy (21.2~eV) ARPES data were acquired using a laboratory-based system consisting of a Scienta R8000 electron analyzer and a Gammadata helium UV lamp.  The angular resolution was  $\sim$ 0.1 $^{\circ}$ and 1 $^{\circ}$ along and perpendicular to the direction of the analyzer slits, respectively. The energy resolution was set at 5 meV. Low photon energy data (6.7~eV) was acquired using tunable laser ARPES system that consists of picosecond Ti:Sapphire oscillator, a fourth harmonic generator\cite{jiang2014tunable} and Scienta DA30 analyzer. The angular resolution was $\sim$ 0.1~$^{\circ}$ and 1~$^{\circ}$ along and perpendicular to the direction of the analyzer slits, respectively. The energy resolution was set at 1~meV. In both cases the energy corresponding to the chemical potential was determined from the Fermi edge of a polycrystalline Cu reference in electrical contact with the sample.  

Band structures have been calculated in density functional theory~\cite{hohenberg1964inhomogeneous_DFT1, kohn1965self_DFT2} (DFT) with PBE~\cite{perdew1996generalized_DFT3} exchange-correlation functional, a plane-wave basis set and projected augmented wave method~\cite{blochl1994projector_DFT4} as implemented in VASP~\cite{kresse1996efficient_DFT5,kresse1996efficiency_DFT6}. For bulk band structures of non-magnetic and block-wise anti-ferromagnetic configurations of La$_{2}$Ni$_{7}$, the primitive hexagonal unit cell with experimental lattice parameters~\cite{levin2004hydrogen_DFT7} have been used with a Monkhorst-Pack~\cite{monkhorst1976special_DFT8} (5 5 3) k-point mesh including the $\Gamma$ point and a kinetic energy cutoff of 270~eV. Tight-binding models based on maximally localized Wannier functions~\cite{marzari1997maximally_DFT9, souza2001maximally_DFT10,marzari2012maximally_DFT11} have been constructed to reproduce closely the bulk band structures in the range of $E_{F}\pm$1~eV with La $sdf$  and Ni $sd$ orbitals. Then the spectral functions and Fermi surface of a semi-infinite La$_{2}$Ni$_{7}$ (001) surface were calculated with the surface Green’s function methods~\cite{lee1981simple_DFT12,lee1981simple_DFT13,sancho1984quick_DFT14, sancho1985highly_DFT15} as implemented in WannierTools~\cite{wu2018wanniertools_DFT16}. 

The five inequivalent Ni sites in the cell are numbered in the following with the Wycoff positions listed in the parenthesis, Ni1 (12k), Ni2 (6h), Ni3 (4f), Ni4 (4e) and Ni5 (2a). Among them, Ni5 (2a) acts as the boundary separating the cell into two blocks along the c-axis with each occupying one of the two prisms in the half-cell blocks. Using the A state with block-wise AFM configuration identified by neutron study \cite{wilde2022}, we have calculated the moment on Ni sites as 0.11 $\mu_{B}$ on Ni1 (12k), 0.27 $\mu_{B}$ on Ni2 (6h), 0.17 $\mu_{B}$ on Ni3 (4f), 0.19 $\mu_{B}$ on Ni4 (4e) and 0.00 $\mu_{B}$ on Ni5 (2a), which are in a good agreement with the earlier DFT calculation\cite{crivello2020relation_La2Ni7_DFTmain}.

\section{Results and discussion}
\begin{figure}[!htb]
\centering
\includegraphics[scale=0.55]{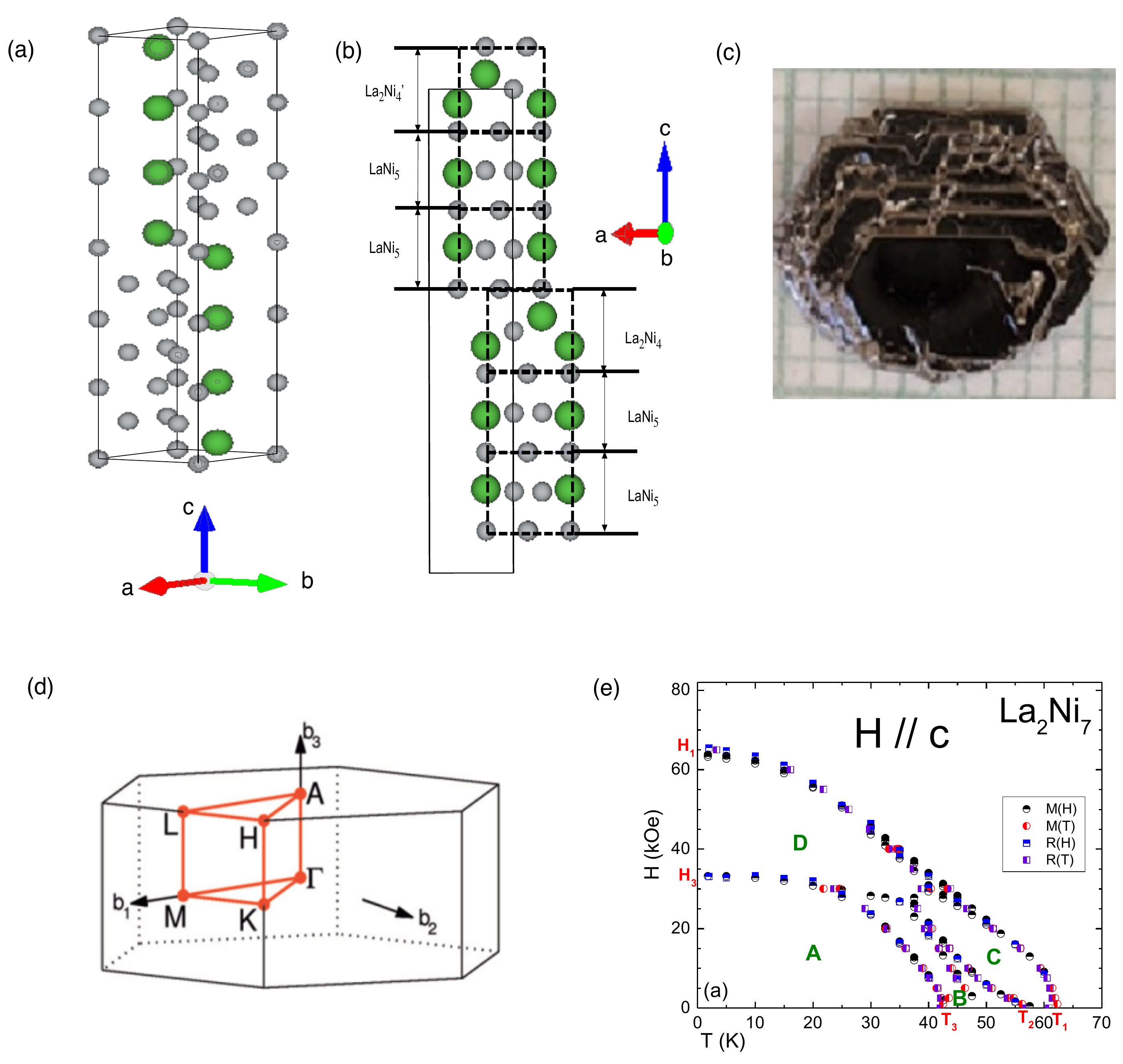}
\caption{Crystal structure, BZ and magnetic phase diagram of La$_{2}$Ni$_{7}$. (a) The crystal structure of La$_{2}$Ni$_{7}$ (space group P6$_{3}$/$mmc$). (b) The crystal structure of La$_{2}$Ni$_{7}$ viewed from different crystallographic orientation. It consists of several La$_{2}$Ni$_{4}$ and LaNi$_{5}$ layers. (c) Picture of single crystal of La$_{2}$Ni$_{7}$. (d) Schematic of the Brillouin zone. (e) H-T diagram of La$_{2}$Ni$_{7}$ from Ref.\cite{ribeiro2021small_La2Ni7_PCC}. }
\label{fig:fig1}
\end{figure}

\noindent
Motivated by the clear delineation of multiple magnetic phases in the H-T phase diagram (based on features in magnetization, specific heat and electrical resistivity data), shown in Fig.~\ref{fig:fig1}(e) \cite{ribeiro2021small_La2Ni7_PCC}, as well as the recent determination of the ordering wave vectors (and their temperature dependencies) for the zero field, A, B, and C phases, we performed the ARPES studies to understand the effects of the magnetic transitions on the electronic structure.
The crystal structure of La$_{2}$Ni$_{7}$ is shown in Fig.~\ref{fig:fig1}(a). Grey spheres represent Ni atoms and green ones are La atoms. The unit cell consists of two [La$_{2}$Ni$_{4}$+2LaNi$_{5}$] blocks as can be seen in Fig.~\ref{fig:fig1}(b) and is highly anisotropic with large values of lattice constant $c=24.71$ \AA \hspace{1pt} compared to the in plane lattice constant $a = 5.058$ \AA. In the lowest temperature, wAFM state ( A state in the H-T phase diagram shown in Fig. \ref{fig:fig1} (d)), each  [La$_{2}$Ni$_{4}$+2LaNi$_{5}$] cluster can be considered as a ferromagnetic (FM) block. Each FM block includes two LaNi$_{5}$ layers and these blocks are separated by the non-magnetic La$_{2}$Ni$_{4}$ layer \cite{crivello2020relation_La2Ni7_DFTmain, wilde2022}. The La$_{2}$Ni$_{4}$ layer at the center of the unit cell provides the inversion center of the nearest neighbors Ni atoms. The top and bottom blocks have opposite orientations of magnetic moments.  As a result, the unit cell consists of two opposing FM blocks with non-magnetic layer in between which result in weakly coupled AFM structure \cite{crivello2020relation_La2Ni7_DFTmain}.  Fig.~\ref{fig:fig1}(c) shows picture of single crystal of La$_{2}$Ni$_{7}$ and Fig.~\ref{fig:fig1}(d) shows the Brillouin zone (BZ) along with the labels of the high symmetry points. Fig.~\ref{fig:fig1}(e) shows H-T phase diagram of La$_{2}$Ni$_{7}$.

%Resistivity as a function of temperature is plotted in Fig. \ref{fig:fig1}(e) and at the high temperatures exhibits  typical metallic behaviour. The inset shows resistivity data at the lower temperatures, where several kinks are visible that may be related to loss of spin disorder. To clarify the origin of these features, we plot the temperature evolution of $\frac{\partial \rho}{\partial T}$ which is shown in Fig. \ref{fig:fig1}(f). There are jumps in the derivative at 42~K, 57~K, and 62~K corresponding to the magnetic phase transitions at zero filed shown in Fig. \ref{fig:fig1}(d).

Motivated by the clear delineation of multiple magnetic phases in the H-T phase diagram (based on features in magnetization, specific heat and electrical resistivity data)\cite{ribeiro2021small_La2Ni7_PCC} and results of recent neutron scattering \cite{wilde2022}, we performed ARPES studies to understand the effects of the magnetic transitions on the electronic structure.  

\begin{figure}
    \centering
    \includegraphics[scale=0.6]{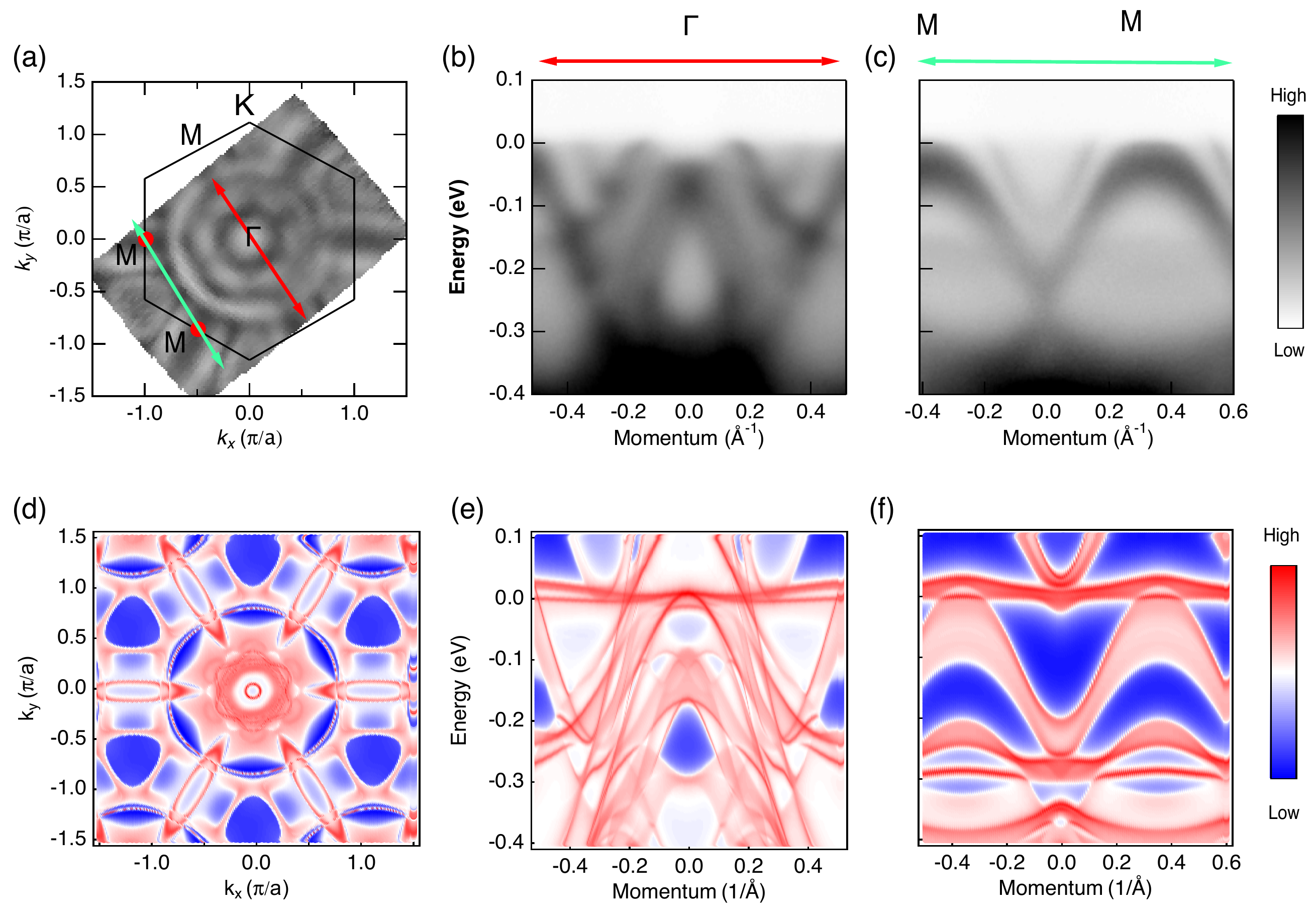}
    \caption{Measured and calculated Fermi surface and band dispersion in PM state. (a) Plot of ARPES intensity integrated within 10~meV of chemical potential that shows the Fermi surface. Red dots represent $M$ point. (b) band dispersion along the $M$-$\Gamma$-$M$ direction. (c) band dispersion along the  $M$-$M$ direction. All measurements performed at T=100~K anbd using 22~eV photons. (d)-(f)  DFT calculations corresponding to data in panels a-c.}
    \label{fig:fig2}
\end{figure}

In Fig.~\ref{fig:fig2}, we focus on Fermi surface (FS) and band dispersion in the  PM state. Fig.~\ref{fig:fig2}(a) shows the ARPES intensity integrated within 10~meV about the chemical potential. Areas of high intensity mark the locations of the FS in the momentum space. The shape of FS reflects six fold symmetry of the crystalline lattice. The FS consists of several hole pockets around the $\Gamma$ and $M$ points. Fig.~\ref{fig:fig2}(b) shows band dispersion along the red solid line ($M$-$\Gamma$-$M$) in panel (a), where one hole-like band and two electron pockets are easily identified. Fig.~\ref{fig:fig2}(c) shows band dispersion along the green solid line ($M-M$) in panel (a). Clearly, we can identify the hole like band dispersion centered at the $M$ point. To elucidate the band structure, we performed DFT calculations. Figs~\ref{fig:fig2} (d)-(f) show the calculated FS and band dispersion based solely on the Ce$_{2}$Ni$_{7}$ type structure in the paramagnetic or nonmagnetic state (i.e. not using any additional magnetic ordering wave-vector). 
Fig.~\ref{fig:fig2}(d) shows calculated FS. The calculation result clearly shows six-fold symmetry, which is consistent with the ARPES result shown in panel (a). Fig.~\ref{fig:fig2}(e) shows calculated band dispersion along the $M$-$\Gamma$-$M$ direction with  visible flat band near the $E_{F}$ and hole-like band dispersion centered at the $\Gamma$ point. Overall, DFT reproduces the shape and locations of the hole pockets and electron pockets reasonably well. Fig.~\ref{fig:fig2}(f)  shows results of the DFT calculation along the $M-M$ direction. The flat band and its hybridization are not observed in the experimental data, quite possibly because they are located somewhat above $E_{F}$. Please note that the experimental value of $E_{F}$ is likely lower by $\sim$ 0.1 eV than in DFT calculation. The energy overlap between the two bands is also greater in the experimental data, whereas the calculations predict complete separation in energy of the two bands by the flat band.

\begin{figure}
\centering
\includegraphics[scale=0.5]{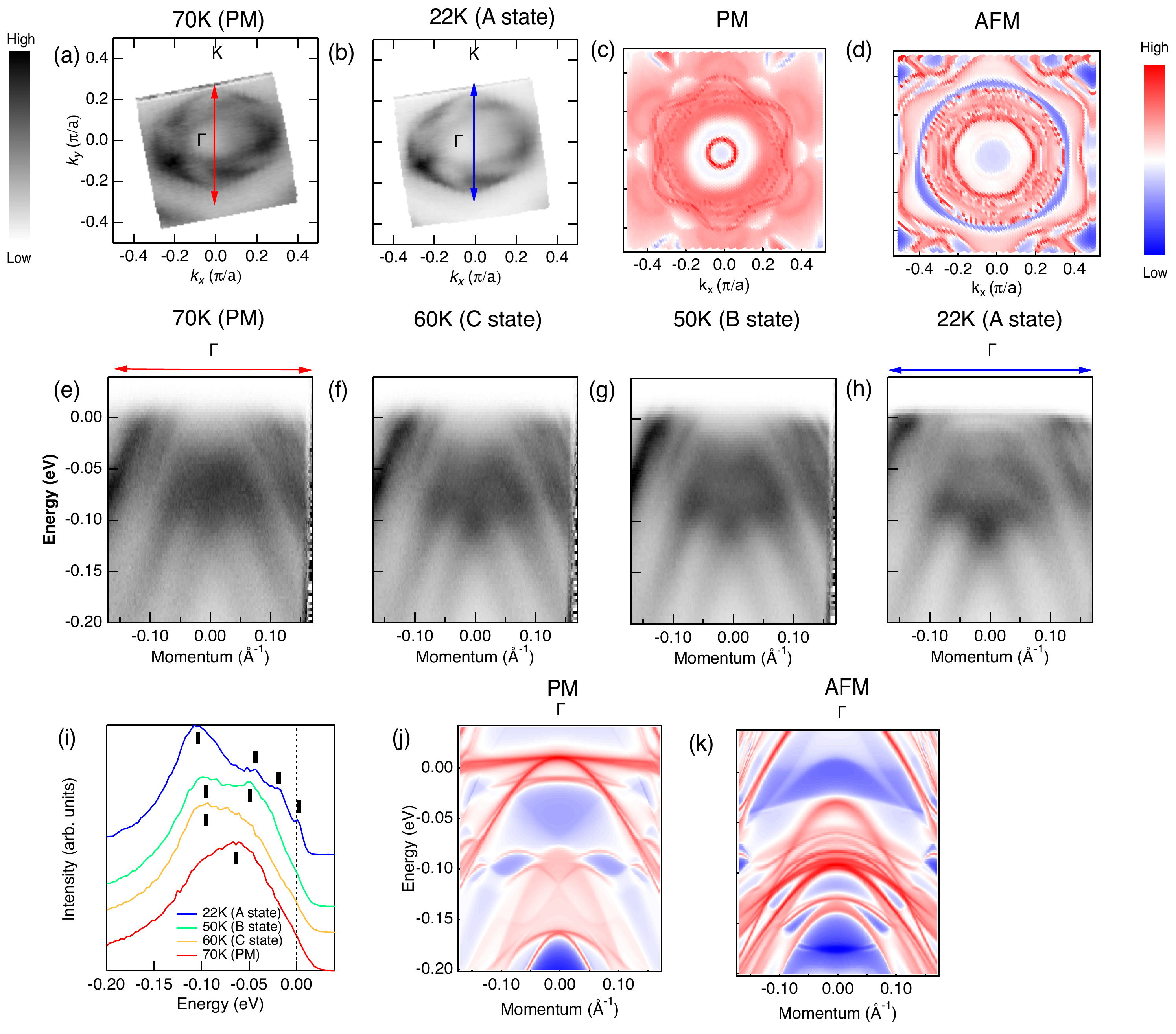}
\caption{ Measured and calculated Fermi surface and band dispersion of La$_{2}$Ni$_{7}$ along the $K$-$\Gamma$-$K$. (a) and (b), plot of the ARPES intensity integrated within 10~meV of the chemical potential in 70~K (PM) and 22~K (A states) respectively. (c) Calculated FS (magnified central portion of Fig.~2d) in the PM state  and (d) AFM state. (e)-(h) Temperature evolution of band dispersion along the $K$-$\Gamma$-$K$ direction. (i) EDCs at the $\Gamma$ point measured PM and each of the magnetic states. Black bars represent the peak positions for each state. (j) Calculated band dispersion in the PM state and (k) State A along the $K$-$\Gamma$-$K$ direction.}
\label{fig:fig3}
\end{figure}

\begin{figure}
\centering
\includegraphics[scale=0.55]{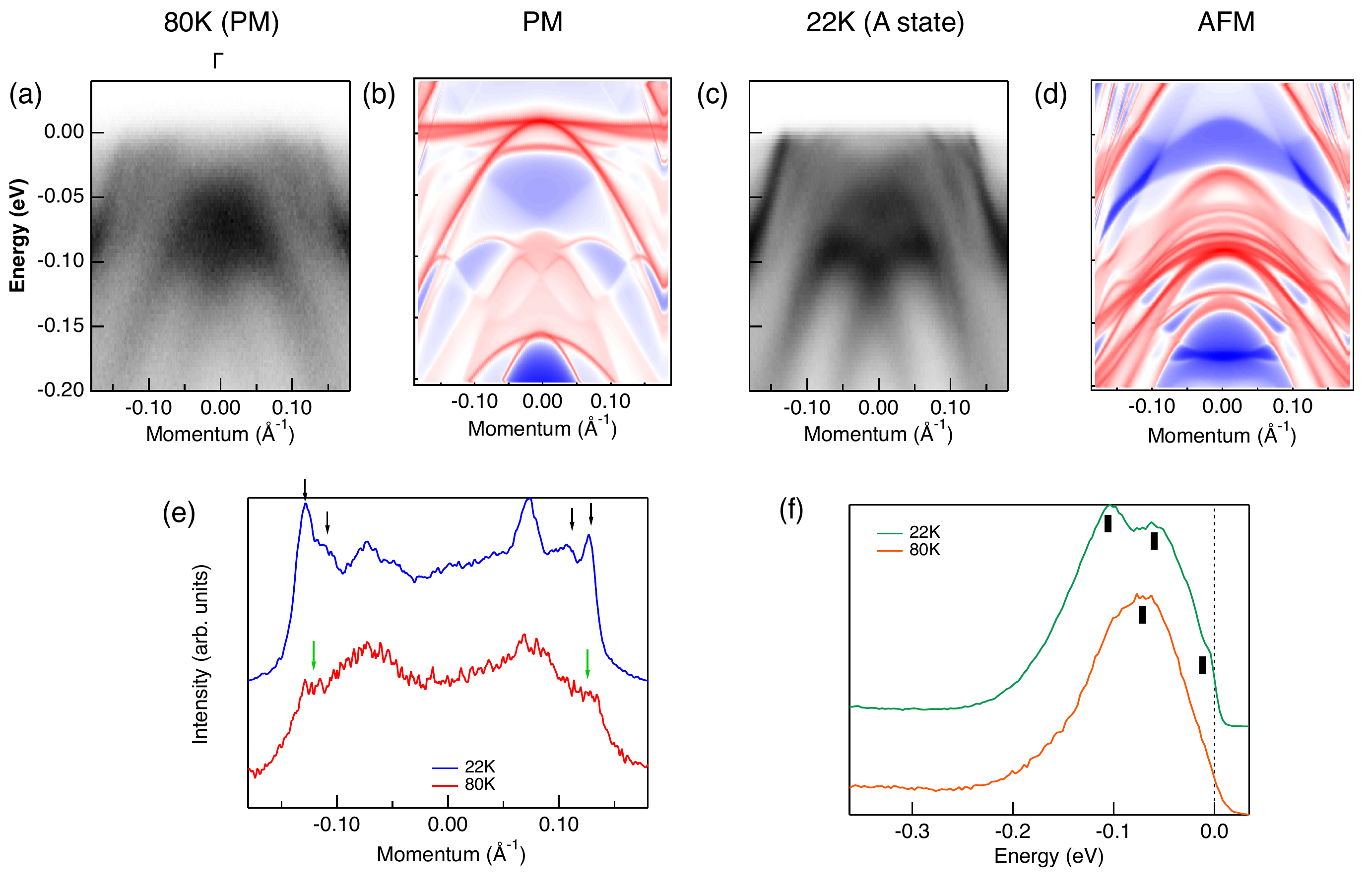}
\caption{Measured and calculated band dispersion of La$_{2}$Ni$_{7}$ along the $\Gamma$-$M$ direction. (a) Band dispersion in PM state (80K) and (b) corresponding DFT calculations of the PM state. (c) The band dispersion of the A state. (d) Calculated band dispersion corresponding to the A state. (e) MDCs at the chemical potential. Black and green arrows mark different peak positions for each temperature. (f) EDCs at the $\Gamma$ point. Black bars mark peak positions. Data acquired using 6.7~eV photon energy.}
\label{fig:fig4}
\end{figure}

We investigated the temperature dependence of the band dispersion. High resolution, temperature dependent data are presented in Fig.~\ref{fig:fig3} and Fig.~\ref{fig:fig4}. Since long-range magnetic ordering introduces gap opening or band splitting at high symmetry points, we investigated the band structure along different high symmetry lines. First, The FS and band dispersion along the $K$-$\Gamma$-$K$ direction are shown in Fig.~\ref{fig:fig3}.  Fig.~\ref{fig:fig3}(a) shows the ARPES intensity integrated within 10~meV about the chemical potential, which reveals the FS at 70~K (PM state) near the $\Gamma$ point. Fig.~\ref{fig:fig3}(b) shows the measured FS for T=22K, i.e. in the T $<$ T$_{3}$, or A, magnetically ordered state.  Fig.~\ref{fig:fig3}(c) shows calculated FS in the PM state near the $\Gamma$ point and Fig.~\ref{fig:fig3}(d) shows the calculated  FS in the AFM state. Whereas the calculation results show significant changes of the FS between PM and AFM states, such changes are not obvious in the measured FS. This discrepancy between experimental and calculation results may be related to the fact that itinerant electrons possess very small magnetic moment. 

Figs.~\ref{fig:fig3}(e)–(h) illustrate the temperature evolution of the band dispersion along the $K$-$\Gamma-K$ direction.  In these panels, we present the band dispersion for the four possible, zero-field, magnetic states shown in the recently determined H-T phase diagram (Fig. \ref{fig:fig1}(d)) Band dispersion in the PM state and C state are shown in Figs.~\ref{fig:fig3}(e) and (f), respectively.   Fig.~\ref{fig:fig3}(g) shows the band structure in the B state. 
The B state data differ from the C and PM phases by prominent splitting of the hole band centered at the $\Gamma$. Band dispersion in the A state is shown in Fig.~\ref{fig:fig3}(h). Upon transition to A phase additional splitting of the hole band at $\Gamma$ occurs.
To understand the details of the band splitting, we plot EDCs at the $\Gamma$ point across the successive magnetic transitions. The result is shown in Fig.~\ref{fig:fig3}(i). The red curve (PM state) shows the highest intensity point  at $\sim$ 50~meV  with the shoulder peak at $\sim$ 100~meV below the $E_{F}$. On the other hand, the yellow curve (C state) exhibits the highest intensity peak at $\sim$ 110~meV and the shoulder peak at $\sim$ 50~meV below the $E_{F}$. The green curve (B state) shows clearly two separated peaks located at 50meV and 110 meV below the $E_{F}$. In addition, the blue curve (A state) shows the emergence of two extra peaks near the $E_{F}$. Consequently, our detailed analysis of the ARPES data suggests that each of the different magnetically ordered states  significantly modifies the band dispersion. We can use calculations to model the band structure for the two states, the paramagnetic state (with no long range magnetic order) and the lowest temperature wAFM state (where we use the magnetic order suggested in Ref \cite{crivello2020relation_La2Ni7_DFTmain, wilde2022})
with results shown in Figs.~\ref{fig:fig3}(j) and (k). Fig.~\ref{fig:fig3}(j) shows calculated band dispersion along the $K$-$\Gamma$-$K$ direction in the PM state. Indicating presence of a flat band and hole pockets near the $E_{F}$. Overall, hole-like band dispersion observed by ARPES measurement in the PM state are described by the calculation reasonably well. The calculation result in the AFM state is shown in Fig.~\ref{fig:fig3}(k). The flat band disappears, and clearly, there are several  band splitting  occurring along the energy axis. This result has a reasonable agreement with our ARPES result shown in the panel (h) (A state) modulo small shift in $E_{F}$. Although  we did not observe significant reconstruction of FS, the band dispersion measured by the ARPES experiment indicates that there are several band splitting below the Fermi level, which is closely related to the magnetic ordering. The increasing number of magnetic band splittings around the Gamma point with decreasing temperature can be related to the multiple non-equivalent Ni sites in these wAFM phases. At the lowest-T A state as calculated from DFT, there are five different Ni sites with also different sizes of magnetic moments, which correspond to different exchange splitting in the bands. For the B and C states with incommensurate wave vector and longer periodicity along the c-axis, the number of non-equivalent Ni sites is even larger. Together with the tilting of the moment in different direction approaching to the paramagnetic phase, the exchange splitting of the bands from these many different site contributions is likely more smeared out than the lowest-T A state as observed in ARPES. 

In order to further study the difference in the dispersion between the PM and wAFM state, given that long range magnetic ordering may induce the band splitting along different high symmetry directions, we investigated band dispersion along different high symmetry direction.  In Fig.~\ref{fig:fig4}, we focus on the band dispersion along the ($M$-$\Gamma$-$M$) direction.  The band structure in the PM state consists of several hole pockets near the $\Gamma$ point (Fig.~\ref{fig:fig4}(a)).  Fig.~\ref{fig:fig4}(b) shows calculated band dispersion along the same cut. In both directions ($M$-$\Gamma$-$M$ and $K$-$\Gamma$-$K$), our ARPES data show broad hole-like band dispersion near the $\Gamma$ point, which may indicate that thermal effect likely play a role in broadening of the signal. Overall, the agreement between data and calculation is reasonable. On the other hand, band dispersion along the different high symmetry directions clearly show different response in magnetically ordered states. Fig.~\ref{fig:fig4}(c) shows band dispersion in the A state. In addition to the band splitting observed at the $K$-$\Gamma$-$K$ direction, our ARPES data along the $M$-$\Gamma$-$M$ direction exhibit additional band splitting near the $E_{F}$ ($k_{\parallel} \sim \pm0.13$ \AA$^{-1}$).  Since 3$d$ orbital from Ni lies close to the $E_{F}$, the splitting may be connected to spin fluctuation from the 3$d$ electrons, but the exact mechanism of the splitting should be scrutinized in future studies. Fig.~\ref{fig:fig4}(d) shows calculated band structure in the AFM state along the $M$-$\Gamma$-$M$ direction, which reproduces reasonably well the band splitting near the $E_{F}$ in the A state and hole pockets at the $\Gamma$ point. 

To understand the details of the magnetic states, we plot MDCs at the $E_{F}$ in Fig.~\ref{fig:fig4}(e).  The red curve shows the MDC in the PM state. The green arrow marks the location of the peak position at 80~K.  The blue curve represents the MDC in the A state.  In both cases, there are clear peaks at $\sim \pm0.7 $~\AA$^{-1}$. Black arrows mark two split peak positions in the A state. Although our calculation model suggests that magnetic ordering emerges along the out of plane direction, the comparison of the MDCs suggests that there is a band splitting along the in-plane direction near the $E_{F}$ in the A state. This may suggest that their is extra magnetic ordering along the in-plane direction. The EDC measured at the $\Gamma$ confirms that the splitting also occurs along energy axis (Fig. \ref{fig:fig4}(f)). Peak positions are marked by black bars. In the A state, we observed band splitting and an extra peak at the $E_{F}$ as compared to the $K$-$\Gamma$-$K$ direction, even though the band splitting is more pronounced in the $K$-$\Gamma$-$K$ direction.

As shown in Fig.3 and 4, the main change near the $\Gamma$ point going from PM to wAFM is the band splitting with lowering temperature. This feature dominates in both the $K$-$\Gamma$-$K$ and $M$-$\Gamma$-$M$ directions. The similarity of band structures between the $K$-$\Gamma$-$K$ and $M$-$\Gamma$-$M$ directions within the $\pm$0.2 ($\frac{\pi}{a}$) range in the DFT calculations agrees well with the ARPES data. This shows that the warping on band structure near the $\Gamma$ point in La$_{2}$Ni$_{7}$ is very small, which can also be seen in the roundish features in the 2D Fermi surfaces within the $\pm$0.2 ($\frac{\pi}{a}$) range in Fig.\ref{fig:fig3}(a)-(d). Such small warping behavior is also consistent with the preferred magnetic moment direction being along the c-axis. (Otherwise strong in-plane anisotropy should be expected if the magnetic moments prefer to align in-plane.)

\section{Conclusions}
In summary, we studied the electronic structure of wAFM La$_{2}$Ni$_{7}$ by means of ARPES and DFT calculations. Our data reveal complex array of band splittings and shifts caused by successive magnetic transitions. While there is some overall agreement in the shape of the FS between calculation and ARPES data in PM state, there are more significant differences in details of the band dispersion. The experiment does not detect the flat band located near $E_{F}$ that was predicted by DFT, possibly because the calculated value of $E_{F}$ is higher than one measured in experiments and this band being it is cut-off by the Fermi function in experimental data. Upon transition to AFM ground state, the flat band is absent in the calculations, in better agreement with data, however the predicted significant changes in the FS are not confirmed by the experiment. The evolution of pattern of band splittings that is observed in experiment upon successive magnetic transitions is not completely reproduced by the calculations. We therefore demonstrate that the series of magnetic transitions in this material leads to significant and rather complex changes in the band structure that is a reasonable agreement with calculations for A state. Since the structural models for remaining states B and C are not yet known, we are not able to perform proper DFT calculations for these.  It appears that La$_{2}$Ni$_{7}$ is a good test system that can be used for improving our understanding of relation between magnetic ordering and band structure and computational modeling.

\clearpage

\clearpage

\bibliography{apssamp}% Produces the bibliography via BibTeX.

\end{document}